\documentclass{osa-article}

\journal{osajournal}

\articletype{Research Article}

\usepackage{lineno}
\linenumbers
\usepackage{float}
\usepackage{amsfonts}
\usepackage[position=t,singlelinecheck=off]{subfig}
\usepackage{graphicx, caption}
\usepackage{hyperref}
\usepackage{cite}

\begin{document}
\nolinenumbers
\title{40 km Fiber Transmission of Squeezed Light Measured with a Real Local Oscillator}

\author{I. Suleiman,\authormark{1,1}, J. A. H.  Nielsen,\authormark{1, 2}, X. Guo,\authormark{1, 3}, 
        N. Jain,\authormark{1, 4}, J. S. Neergaard-Nielsen,\authormark{1, 5}, T. Gehring,\authormark{1, *} and U. L. Andersen,\authormark{1, $\dagger$}}

\address{\authormark{1}Center for Macroscopic Quantum States (bigQ), Department of Physics, Technical University of Denmark, 2800 Kongens Lyngby, Denmark}

\email{\authormark{1} isule@fysik.dtu.dk, \authormark{2} jearn@fysik.dtu.dk, \authormark{3} xueshiguo@tju.edu.cn, \authormark{4} Nitin.Jain@fysik.dtu.dk, \authormark{5} jsne@fysik.dtu.dk, \authormark{*} tobias.gehring@fysik.dtu.dk, \authormark{$\dagger$} ulrik.andersen@fysik.dtu.dk} 



\begin{abstract}
We demonstrate the generation, 40 km fiber transmission, and homodyne detection of single-mode squeezed states of light at 1550 nm using real-time phase control of a locally generated local oscillator, often called a ''real local oscillator'' or ''local local oscillator''. The system was able to stably measure up to around 3.7 dB of noise suppression with a phase noise uncertainty of around 2.5$^\circ$, using only standard telecom-compatible components and a field-programmable gate array (FPGA). The compactness, low degree of complexity and efficacy of the implemented scheme makes it a relevant candidate for long distance quantum communication in future photonic quantum networks. 
\end{abstract}

\section{Introduction}
\label{Introduction}
Squeezed states of light have by now been generated for more than three decades \cite{andersen_30_2016} and have become a ubiquitous resource in optical quantum information science.  
Unlike coherent states of light, squeezed light has the outstanding property of exhibiting lower noise uncertainty than the fundamental shot noise in some of its quadratures -- known as squeezing -- while conjugated quadratures exhibit uncertainties above the shot noise limit -- known as anti-squeezing \cite{yuen_two-photon_1976, caves_quantum-mechanical_1981}. This fundamental quantum property of squeezed states has been the engine of numerous quantum sensing experiments, such as the quantum-enhanced measurements of gravitational waves~\cite{aasi_enhanced_2013} and vibrational modes of molecules~\cite{de_andrade_quantum-enhanced_2020,casacio_quantum-enhanced_2021}, and recent quantum computing models 
including Gaussian boson sampling \cite{arrazola_quantum_2021,zhong_phase-programmable_2021} and measurement-based quantum computing \cite{larsen_deterministic_2019,asavanant_generation_2019,larsen_deterministic_2021}. 
Furthermore, the use of squeezed states (compared to coherent states) improves the performance of continuous variable quantum key distribution systems by increasing the tolerance to optical loss, excess noise and imperfect error correction~\cite{madsen_continuous_2012,gehring_implementation_2015,jacobsen_complete_2018}, effectively allowing longer transmission distances or improved security. 

The most common technique employed to measure squeezed states of light is homodyne detection, which implements the ideal measurement of one quadrature of the electromagnetic field by using a strong reference beam, called the local oscillator, in conjunction with a (balanced) detection system~\cite{yuen_two-photon_1976,lvovsky_continuous-variable_2009}.
Because of its versatility, simplicity and high efficiency, it is a widely employed measurement technique in quantum information protocols including the recent demonstrations of Heisenberg limited quantum sensing~\cite{nielsen_deterministic_2021}, measurement-induced quantum computing gates~\cite{larsen_deterministic_2021}, long-distance quantum key distribution~\cite{zhang_long-distance_2020} and high-speed quantum random number generation~\cite{gehring_homodyne-based_2021}.

In most quantum optical proof-of-concept experiments, the homodyne detection process is carried out with a local oscillator that originates from the same laser as the one that produces the signal, e.g. the squeezed state. The advantage in doing so is that the local oscillator is naturally frequency-synchronized with the signal, and the remaining slow phase fluctuations can be easily stabilized using simple control systems. However, in real-life applications on quantum communication, distributed quantum sensing and networked quantum computing where the information-carrying optical signal, e.g. the squeezed state, needs to be transmitted through optical fibers to remote locations, a centrally distributed local oscillator is problematic for several reasons. Firstly, the optical power of the local oscillator decreases exponentially with propagation distance, and therefore, to provide sufficient optical power at the various nodes in the network for enabling low-noise homodyne detection, a significant amount of power must be distributed through the fiber network. In addition to this direct power wastage, the massive power distribution might also lead to deteriorating non-linear optical effects (such as Brillouin scattering) and to distortion of the quantum state of the signal \cite{qin_transferring_2019, yaman_guided_2021}. 
Secondly, transmitting the local oscillator along with the signal is also fundamentally problematic for continuous-variable quantum key distribution (CVQKD), as there are several eavesdropping strategies that explicitly make use of the transmitted local oscillator to corrupt the coherent detection at the trusted receiver, therefore breaching the security of the QKD protocol \cite{haseler_testing_2008,jouguet_experimental_2013,ma_local_2013, huang_quantum_2013}. 

Hence, it is preferable to generate the local oscillator locally at the receiving nodes. Obviously, real-time control of the phase difference between the signal and the local oscillator becomes more challenging as the two laser sources are in general not frequency-locked; partly for this reason, many current proof-of-concept CVQKD systems, while using a real local oscillator, use heterodyne reception and do not stabilize the phase in real-time, but rather rely on phase estimation and compensation in post-processing, possibly invoking elaborate phase estimation techniques \cite{zhao_phase_2019}. Moreover, due to the complications associated with a real local oscillator, all previous experiments on quantum communication with squeezed light have employed a local oscillator derived from the same laser as they used for squeezed light generation.

In this article, we report on the generation, transmission and homodyne detection of single-mode squeezed states of light at the wavelength of 1550 nm using a real local oscillator with real-time phase control. By exploiting a real-time feedback system, we controlled the frequency and phase of an locally-generated  local oscillator to such an extent that it enabled homodyne detection with a phase uncertainty of 2.51$^\circ$ of a squeezed state that had propagated up to 40\,km in a telecom fiber. Our demonstration constitutes a critical milestone in the construction of a quantum information network based on continuous variable quantum systems.        
\section{Methods}
\label{Methods}
\begin{figure}[H]
	\centering
	 		\captionsetup{width=150pt}
	\subfloat[]{
		\includegraphics[scale=.7]{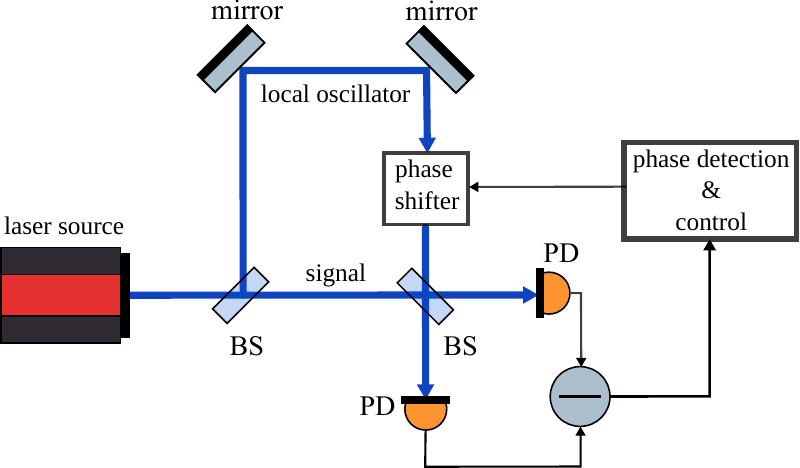}}
	\hspace{1cm}
	\captionsetup{width=140pt}\subfloat[]{
		\includegraphics[scale=.7]{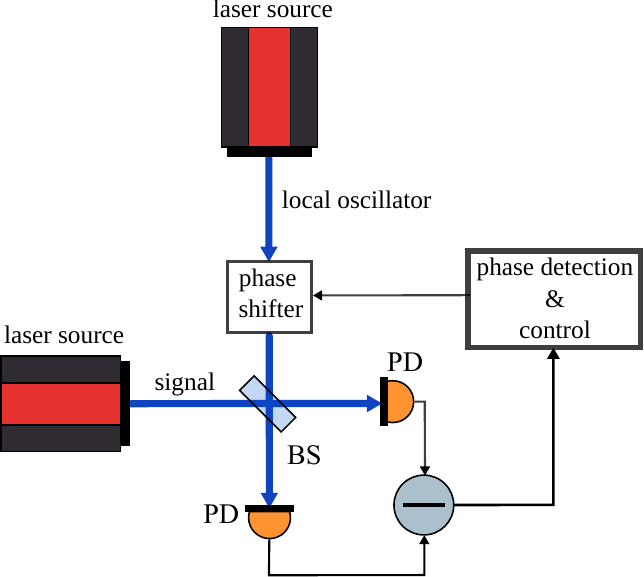}}
	\captionsetup{width=\textwidth} \caption{Conceptual schematic of a typical homodyne detection system. The signal is interfered with the local oscillator onto a balanced beam splitter (\textbf{BS}). The modes output from the BS are input to two photodetectors (\textbf{PD}) and the resulting photocurrents are electronically subtracted, yielding a photocurrent that is proportional to the phase difference between the signal and the local oscillator modes. The latter is used to estimate and control the phase difference, by shifting the phase of the local oscillator relative to the signal (\textbf{phase shifter}). The local oscillator can be (a) generated from the same laser source that produces the signal, or (b) generated from an independent laser source. Blue lines represent optical modes, black lines represent electricals signals.}
	\label{figure:homodyne detection -  concept}
\end{figure}

Balanced homodyne detection works by interfering the mode to be detected, here named the \textit{signal}, with a much more intense optical mode, known as the \textit{local oscillator} (LO), on a balanced beam splitter. The outputs of the beam splitter are simultaneously detected with two photodiodes, and the resulting photocurrents are electronically subtracted to produce a signal which is proportional to a specific quadrature of the signal. The quadrature being measured is governed by the phase difference of the signal and the local oscillator, so in order to enable a long-term measurement of a specific quadrature -- as required for most applications -- the phase relation between the two input modes must be kept constant. This is often realized with an active feedback control system. A conceptual schematic of homodyne detection is shown in figure \ref{figure:homodyne detection -  concept}.

For the detection of squeezed light, it is common to use the same laser for delivering power to the local oscillator and for generating the squeezed state which means that the central frequency component of the squeezed light is naturally synchronized with the local oscillator frequency. Such frequency synchronization is indeed required as the detection of squeezed light corresponds to the precise detection of adjacent, quantum correlated, frequency side bands located symmetrically around the reference frequency. This in-built frequency synchronization also means that small frequency drifts of the laser play no role in the performance of the measurement.  

\begin{figure}
	\centering
	\captionsetup{width=\textwidth}
	\subfloat[]{
		\includegraphics[scale=.15]{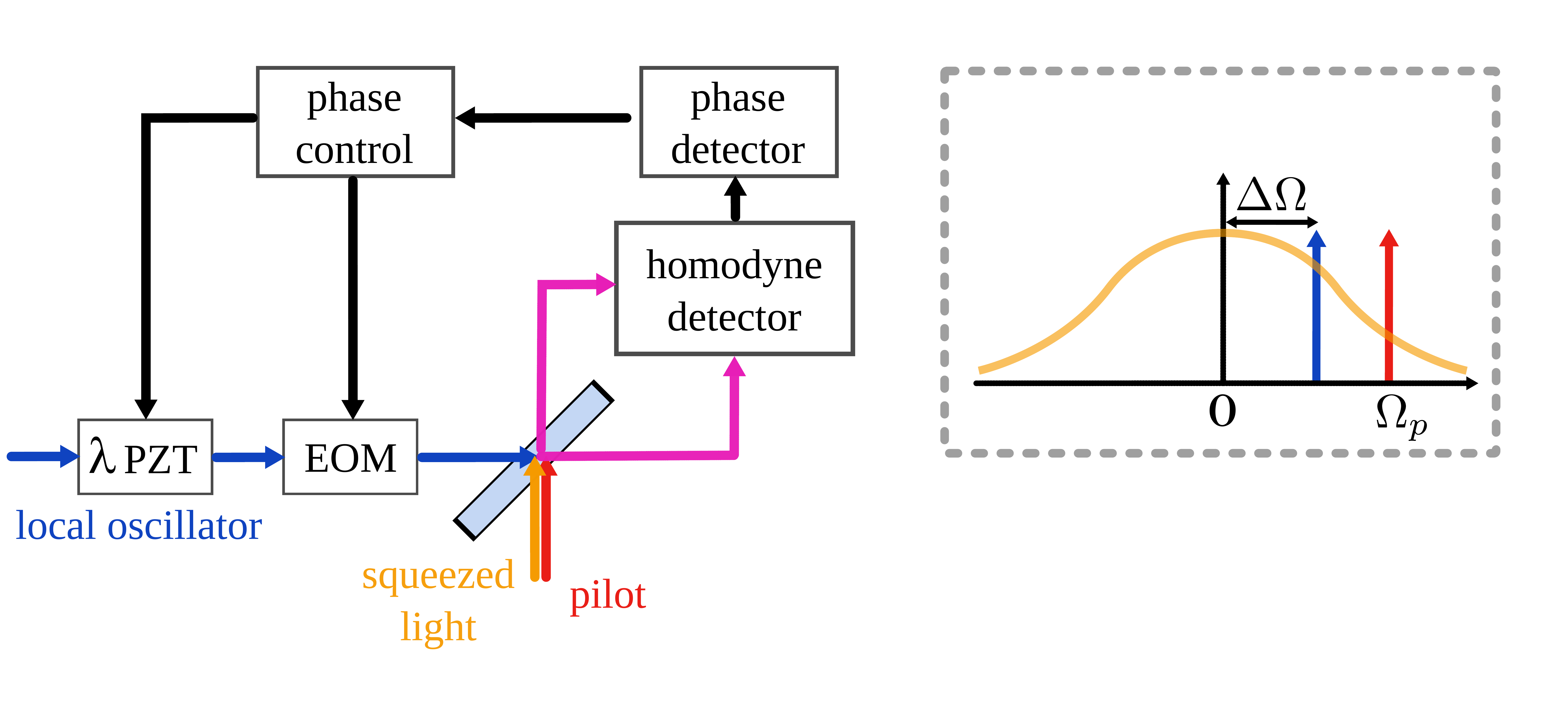}}
	\vspace{0.1pt}
	\centering
	\subfloat[]{
		\includegraphics[scale=.13]{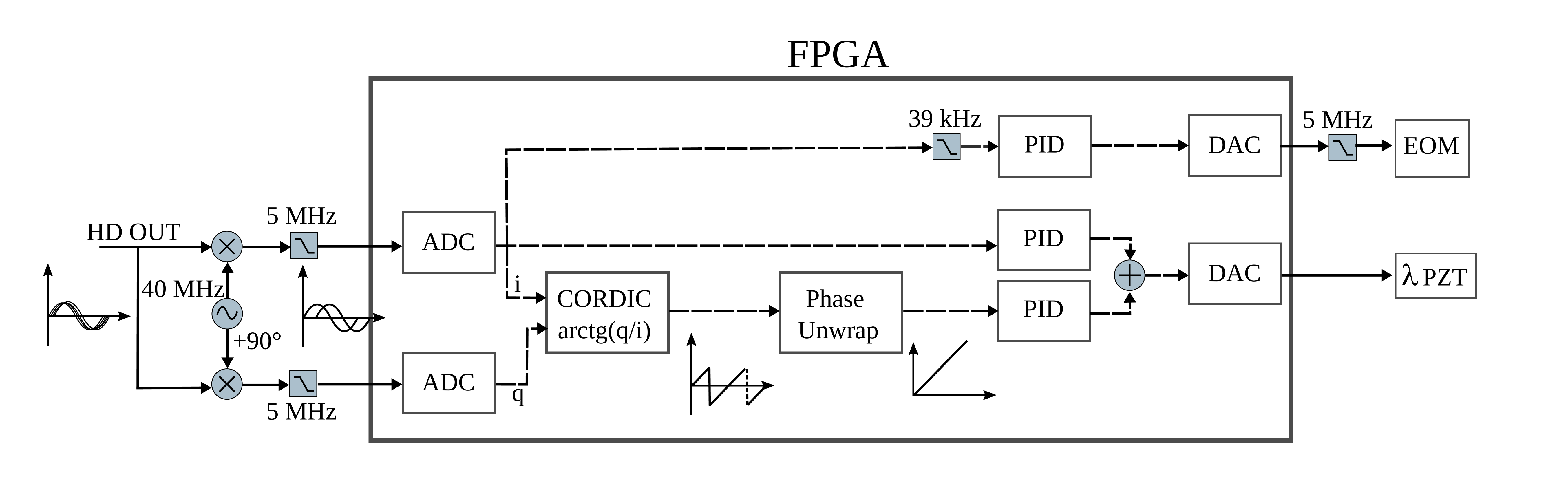}}
	\caption{(a) Conceptual schematic of homodyne detection using a real local oscillator. The frequency-shifted pilot interferes with the real local oscillator on a balanced beam splitter. A phase detector delivers suitable error signals to control the frequency of the local oscillator, via a piezoelectric wavelength modulator ($\lambda$ PZT), and its phase relative to the pilot, through an electro-optic phase modulator (EOM). Wavelength control aims at shifting the local oscillator by $f_p=\Omega_p/2\pi$ = 40 MHz from the pilot to perform homodyne detection of the squeezed mode. The diagram on the right illustrates the situation in the frequency domain, with the local oscillator, pilot and squeezed fields indicated by the same colors as on the left. \\(b) Conceptual schematic of the phase control system for the real local oscillator. The unwrapped instantaneous phase is generated from the pilot/local oscillator interference contained in the signal ''HD OUT'' inside a field gate-programmable gate array (FPGA), via CORDIC algorithm and phase unwrapping, and is used for a coarse control on the frequency of the local oscillator. Finer frequency control is performed using the demodulated interference $i(t)$ at the pilot offset frequency $f_p$. The same error signal is then used for fast phase correction by the electro-optic phase modulator. ADC: analog-to-digital converter. PID: proportional-integral-derivative controller. DAC: digital-to-analog converter.}
	\label{figure:LLO concept}
\end{figure}

To achieve homodyne detection of squeezed light at a fixed phase angle with a distinct local oscillator where frequency synchronization is not inherent, synchronization must be established in real time to counteract the relative frequency drifts. This must be done with a precision that allows for the simultaneous detection of symmetrically correlated sidebands such that the squeezing is revealed. Towards this end, we employ a pilot tone technique as illustrated in Fig.~\ref{figure:LLO concept}a: 
A pilot tone is a coherent excitation which is extracted from the central laser (which is used for squeezed light generation) and frequency shifted with respect to the squeezed light carrier by $\Omega_p=2\pi f_p$. The pilot tone is combined with the squeezed light mode, residing as a single frequency sideband, thereby providing a frequency and phase reference for the local oscillator upon reception at the homodyne detection station. We note that pilot tones are an essential part of the coherent control scheme commonly used to actively control squeezed light sources~\cite{vahlbruch_observation_2008,aasi_enhanced_2013,arnbak_compact_2019} and, thus, they are readily available in many squeezed light experiments. At the receiver station, the frequency offset and phase difference relative to the real local oscillator are measured through interference at a beam splitter as shown in the figure, and the resulting signals are subsequently used to control the frequency and phase of the real local oscillator mode to compensate for any phase and frequency drifts. The feedback signal is divided into a slow port for frequency control of the LO laser and a fast port for phase control via an electro-optics modulator. 

The output signal of a balanced homodyne receiver measuring the pilot tone with an independent laser as LO is proportional to
%
%
\begin{equation}
\sqrt{P_{p} P_{LO}}\cos\left((\Omega_p + \Delta \Omega(t))t + \Delta\phi(t) - \phi_\text{set}\right)\ ,
\label{eq:homodyne AC oscillation}
\end{equation}
where $P_{p}$ and $P_{LO}$ are the optical powers of the pilot and local oscillator at the input to the homodyne detector, $\Delta \Omega(t)$ is the time dependent angular frequency offset of the two lasers, and $\Delta\phi(t)$ is the time dependent phase difference between LO and pilot (modulo their frequency offset). The goal of our active feedback scheme is to achieve $\Delta \Omega(t) = \Delta\phi(t) = 0$ and to thereby perform homodyne detection at the phase angle $\phi_\text{set}$.



The laser we employed as LO laser in the experiment was an NKT Photonics E15 which allows for frequency tuning via a piezo-electric transducer (PZT). We complemented this slow frequency actuator with a fast electro-optical phase modulator (EOM), cf.\ Fig.~\ref{figure:LLO concept}a. The feedback loop, consisting of phase detection and control, was implemented in a field-programmable-gate array (FPGA) as shown in Fig.~\ref{figure:LLO concept}b. To limit the required sampling rate of the analog-to-digital converters (ADCs) connected to the FPGA we first electrically down-mixed the output of the homodyne detector, see Eq.~\ref{eq:homodyne AC oscillation}, with two orthogonal sinusoidal signals with amplitude $A_r$, frequency $\Omega_p$ and phase $\phi_\text{set}$. After down-mixing we low-pass filtered the signal with 5 MHz cutoff frequency. The electrical sinusoidal signals have to be synchronized with the generation of the pilot tone at the squeezed light source which can either be achieved by transmitting a clock signal using a wavelength multiplexed channel or a dedicated servicing fiber, which is a standard technique in quantum communication, or potentially by more sophisticated digital-signal-processing~\cite{chin_machine_2021}. After down-mixing and low-pass filtering, the two signals are given by
\begin{align}
i(t) & \propto \sqrt{P_{p} P_{LO}} A_r \cos\left(\Delta\Omega(t)t + \Delta\phi(t)\right)\ , \\
q(t) & \propto \sqrt{P_{p} P_{LO}} A_r \sin\left(\Delta\Omega(t)t + \Delta\phi(t)\right)\ .
\end{align}
These were sampled by two ADCs with sampling period $T_s$, resulting in a discretization of time as $t=nT_s$ with sample index $n$. 

In the FPGA, we established two signal paths corresponding to the two actuators: A fast path using the EOM as actuator, and a slow path actuating with the PZT. For the PZT actuation path we estimated the instantaneous phase error by calculating the $\arctan$ between $q(t)$ and $i(t)$ using a coordinate rotation digital computer (CORDIC) algorithm\cite{bhukya_design_2021} which yields
\begin{equation}
\Phi_e(nT_s) \simeq \Delta\Omega(nT_s) nT_s + \Delta\phi(nT_s) \pmod{[-\pi, \pi]}\ .
\label{eq:wrapper phase error}
\end{equation}
If the phase error exceeds $[-\pi, \pi]$, $\Phi_e$ undergoes an abrupt transition, often called \textit{phase wrap}. This can be avoided by accumulating phase increments between consecutive samples, namely $\Phi_e(nT_s) - \Phi_e((n-1)T_s)$ and detecting a phase wrap event by comparing the last two bits of the binary representation of the current phase increment, and compensating for it by adding or subtracting $2\pi$ \cite{kumm_digital_2008}. Phase unwrapping allows to detect a wider range of phase fluctuations, at the cost of decreased signal resolution due to a fixed number of bits used in the implementation. 

To generate the actuator signal for the PZT, the output of the phase unwrapping algorithm was injected into a proportional-integral (PI) controller whose output was added to the output of another PI controller taking directly $i(t)$ as error signal. The latter  allowed to control the medium frequency range which was not possible with the instantaneous phase error signal due to latency.

The fast path used $i(t)$ as error signal. This was low-pass filtered at 39 kHz, since at higher frequencies the measured phase fluctuations hit the noise floor of the detection system. After PI control, the signal was converted to the electrical domain and low-pass filtered at 5 MHz to eliminate high frequency electronic noise.  

The lock was acquired by first tuning the frequency of the LO laser into the capture range of our controller via the laser's temperature. The capture range was determined by the 5 MHz lowpass filter before analog-to-digital conversion. The lock was then engaged by first disabling the fast path as well as the medium frequency part of the slow path which allowed us to achieve a frequency lock first. Afterwards the two additional paths were enabled to lock the phase without cycle slips. We note that the homodyne angle can be adjusted by changing the phase $\phi_\text{set}$ of the electrical LO used for downmixing the homodyne detector output. 

\begin{figure}[b]
	\centering
	\includegraphics[scale=.7]{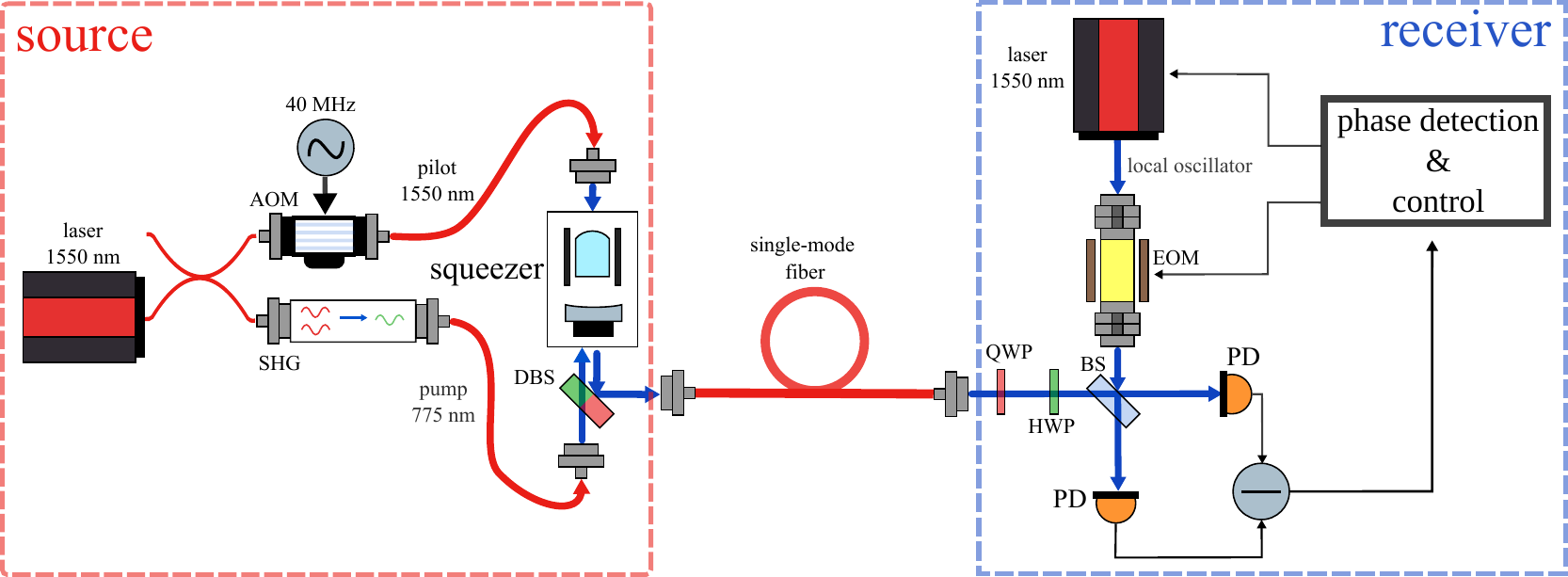}
	\captionsetup{width=\textwidth}\caption{Basic scheme of the experimental system. Red lines represent light propagating in optical fiber. Blue lines represent light propagating in free space. Black lines represent electrical signals. AOM: acousto-optic modulator. SHG: second-harmonic generation module. DBS: dichroic beam splitter. EOM: electro-optic modulator. HWP: half-wave plate. QWP: quarter-wave plate. BS: beam splitter. PD: photodiode. The figure does not include any employed control system.}
	\label{figure:physical setup}
\end{figure}

The described phase locking scheme was experimentally tested with the experimental setup in Fig.~\ref{figure:physical setup} showing a schematic for squeezing generation, fiber transmission and detection.  

The laser that generates the squeezed and the pilot modes is an NKT Koheras ADJUSTIK E15, at 1550.12 nm wavelength.
Squeezed vacuum at 1550\,nm was generated through parametric downconversion using a periodically poled potassium titanyl phosphate (KTP) crystal placed inside an optical cavity resonant for the squeezed field and the pump field at 775\,nm. The pump was generated  a commercial second harmonic generation module (NTT Electronics WH-0775-000-F-8-C), consisting in a lithium niobate (LN) crystal with waveguide. 
While the pump field was used to lock the squeezed light source's cavity on resonance, we used the coherent control scheme for the phase. An acousto-optic modulator (AOM) frequency-shifted a small fraction of the fundamental laser beam at 1550\,nm by $f_p=40$\,MHz which was then injected into the cavity and its phase was locked with respect to the pump field. This constituted the pilot signal used at the detection stage. It had a power of about 5\,$\mu$W leaving the cavity and it copropagated with the squeezed light. More details about the squeezed light source can be found in Ref.~\cite{arnbak_compact_2019}.

Reflected off a dichroic beam splitter (DBS), the squeezed and pilot modes were coupled from free space into a standard single-mode optical fiber (SMF-28) and propagated towards the free space receiver station, where homodyne detection of the squeezed mode took place using a separate laser source at 1550.12\,nm wavelength (NKT Koheras MIKRO E15) as local oscillator. To match the polarization of the squeezed light after transmission through the fiber to the local oscillator we used a quarter waveplate and a half waveplate.


\section{Results}

\begin{figure}
	\centering
	\captionsetup{width=100pt}\subfloat[]{\includegraphics[scale=0.2]{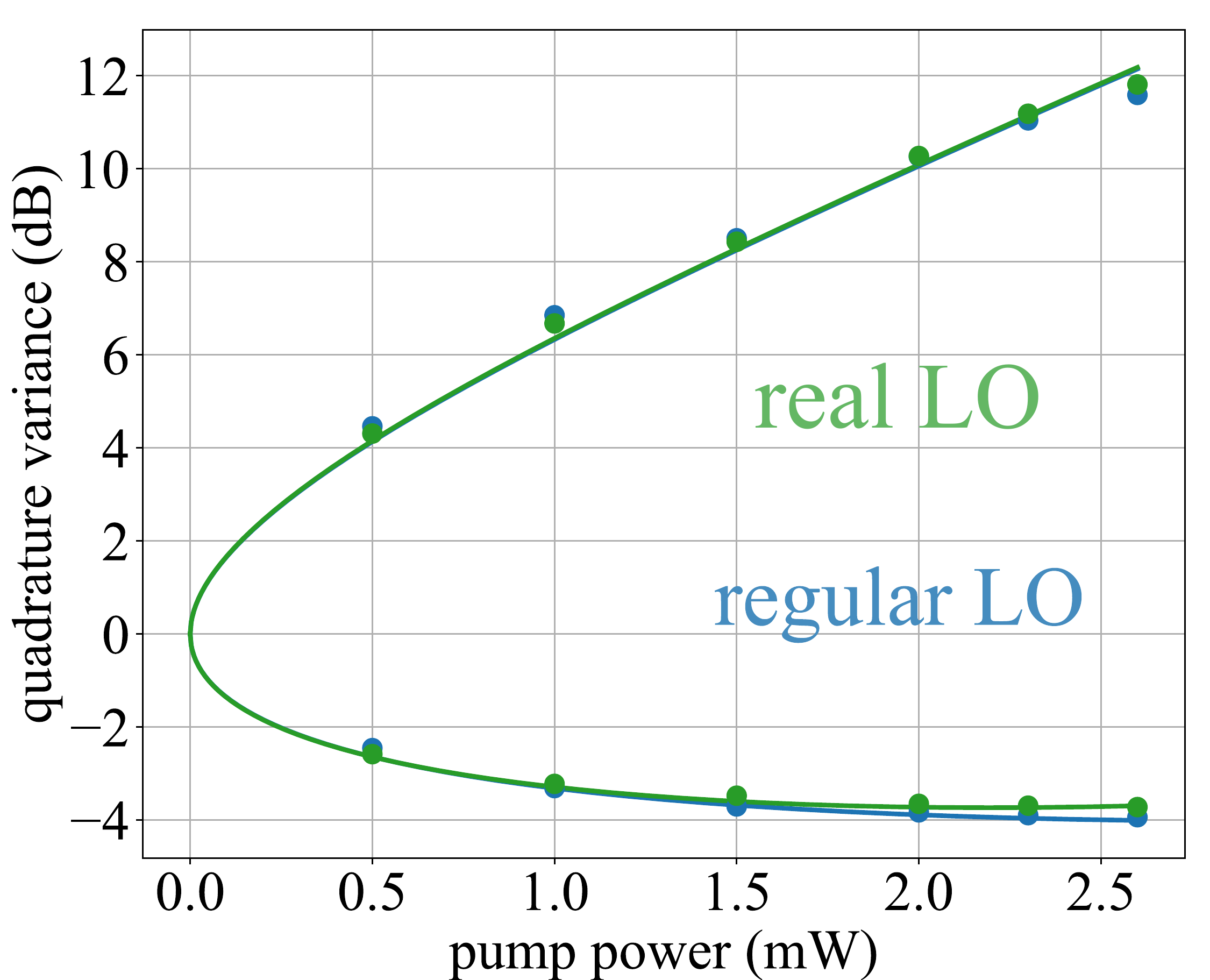}}
	\hspace{10 pt}
	\captionsetup{width=100pt}\subfloat[]{\includegraphics[scale=0.2]{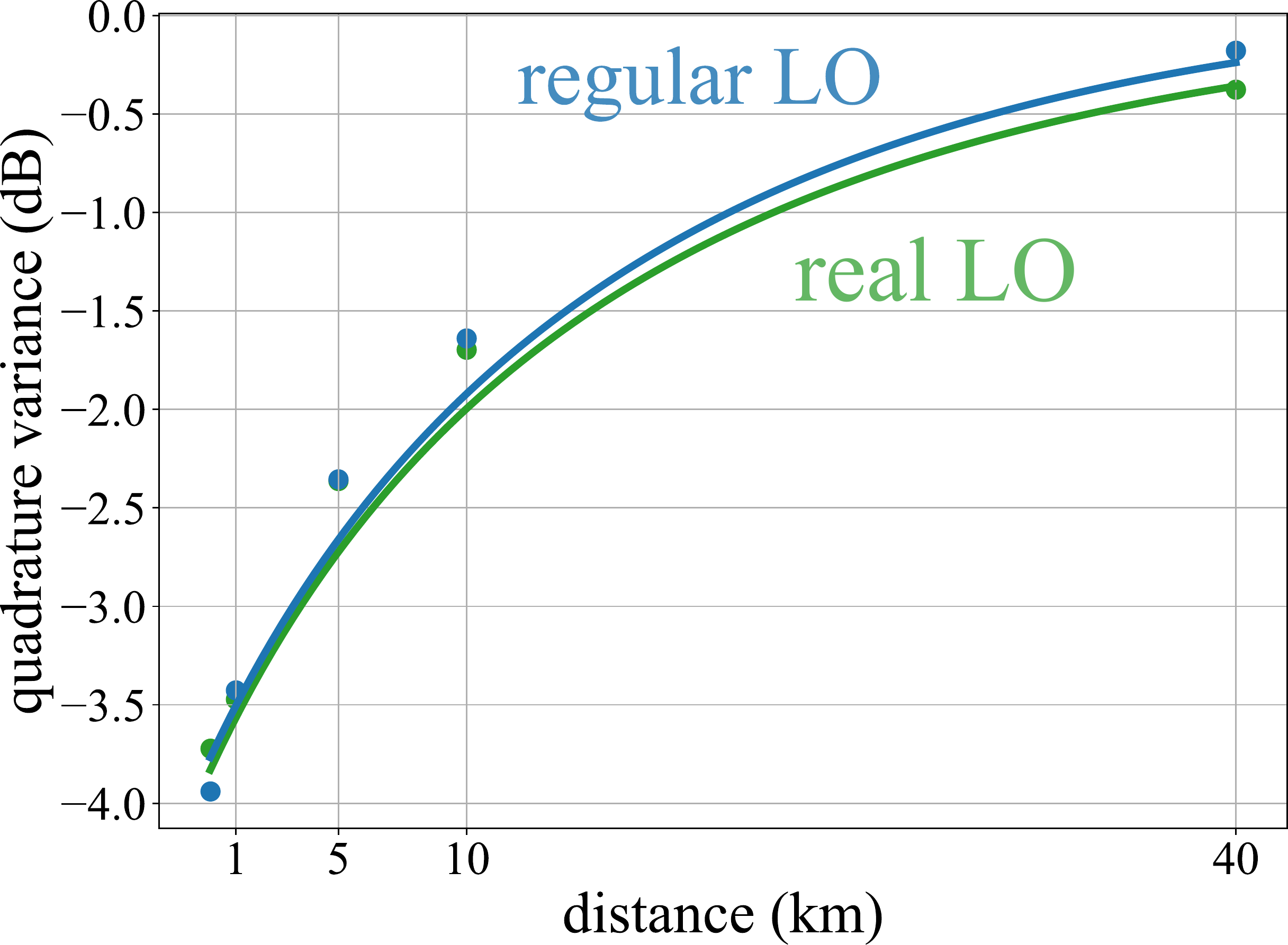}}
	\vspace{2pt}
	\captionsetup{width=100pt}\subfloat[]{\includegraphics[scale=0.2]{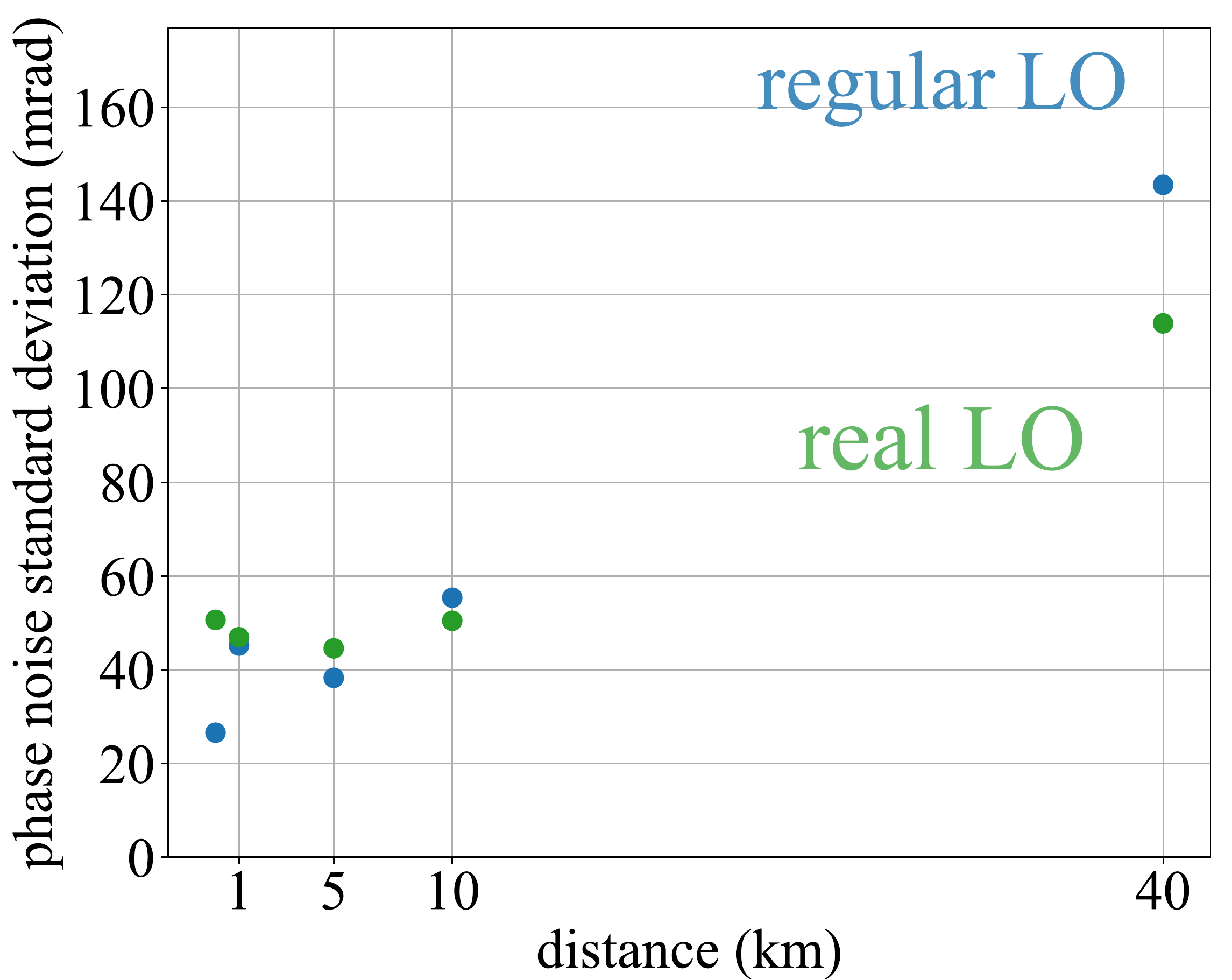}}
	\hspace{10 pt}
	\captionsetup{width=100pt}\subfloat[]{\includegraphics[scale=0.2]{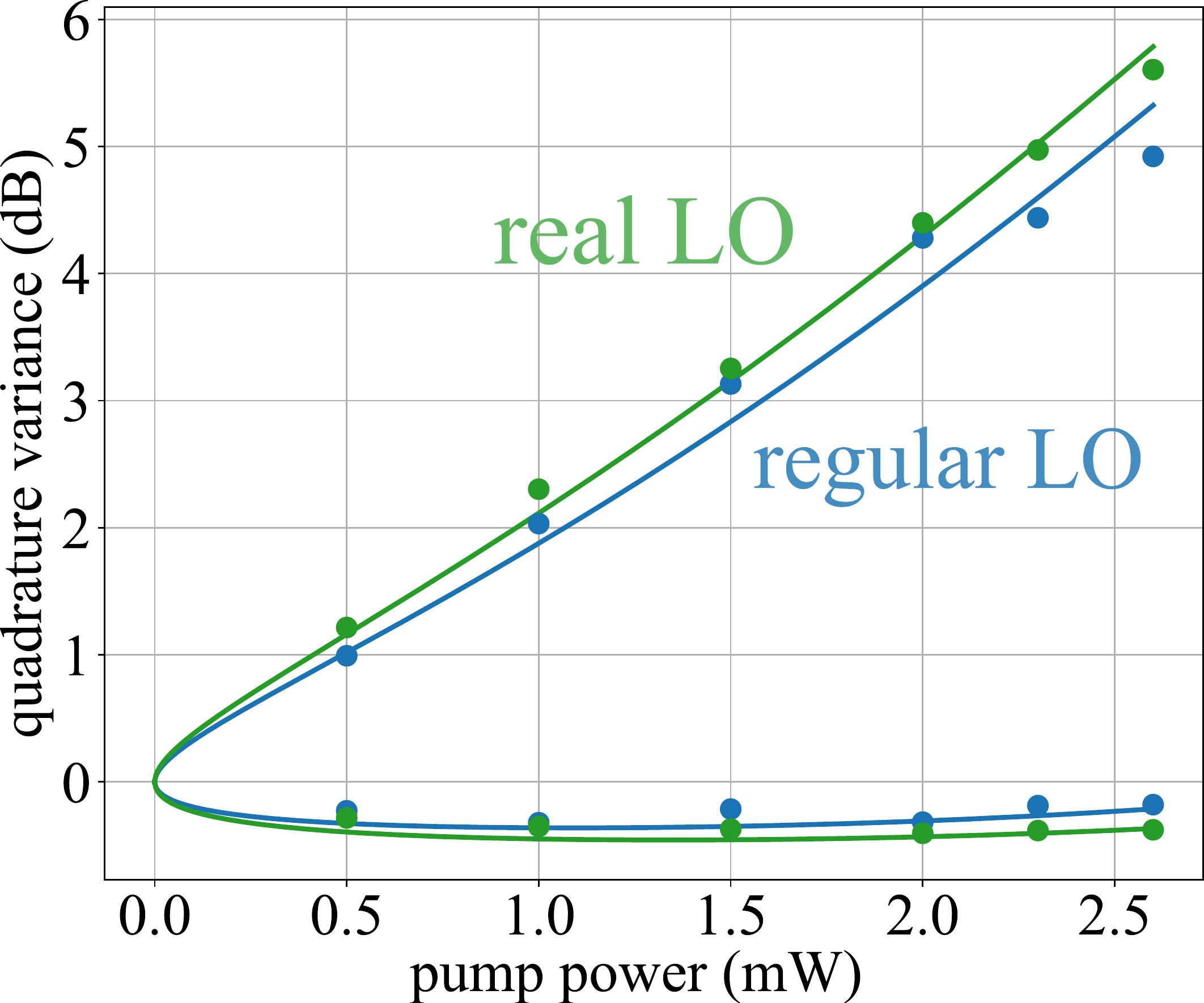}}
	\vspace{2pt}
	\captionsetup{width=\textwidth}
	\caption{Experimental results. (a) and (d) present the measured squeezed (below zero) and antisqueezed (above zero) quadrature variance with the two types of local oscillator, for different pump field optical powers coupled into the squeezer, and fiber lengths of 10 m (figure (a)) and 40 km (figure (d)). Figures (b) and (c) represent the squeezed quadrature variance and estimated phase noise standard deviation, respectively, for 2.6 mW pump field optical power and different fiber lengths. real local oscillator: green. Regular local oscillator: blue. Dots represent experimental data, solid lines represent fitting curves. If not visible, the error bars are smaller than the dot sizes.}
	\label{fig:results global}
\end{figure}

The experimental results are shown in Fig.~\ref{fig:results global}. First, we transmitted the squeezed light through a 10\,m fiber and compared the performance of the phase locked loop to a measurement for which we used a regular LO transmitted along with the squeezed light. This is shown in Fig.~\ref{fig:results global}a where we plot the variance of the squeezed ($V_-$) and anti-squeezed ($V_+$) quadratures for varying pump power of the squeezed light source. To the experimental data points we fitted a model given by~\cite{aoki_squeezing_2006,collett_squeezing_1984}
\begin{equation}
V_\pm = V_\pm^0\ \frac{1+e^{-2\sigma^2}}{2} + V_\mp^0\ \frac{1-e^{-2\sigma^2}}{2}\ ,
\label{eq:quadrature PSD}
\end{equation}
with
\begin{equation}
V_\pm^0 = 1 \pm \frac{4\eta F_g}{(1\mp F_g)^2 + (f/f_\mathrm{sqz})^2}\ .
\end{equation}
Here, $\eta$ is the overall efficiency, $F_g = \sqrt{P_\text{pump}/P_\text{threshold}}$ with $P_\text{pump}$ being the optical power of the pump field coupled into the squeezed light source and $P_\text{threshold} \simeq 5.12$ mW\cite{arnbak_compact_2019} the pump power where the lasing threshold is reached, $f = 12.2$ MHz is the measurement frequency, $f_\text{sqz} \simeq 66$ MHz \cite{arnbak_compact_2019} is the half-width-half-maximum of the squeezed light source's frequency response and $\sigma$ is the standard deviation of the phase noise. $V_\pm^0$ are the variances with zero phase noise.

As can be seen from the figure, the measured squeezing with the phase-locked real LO laser is almost the same as  with the regular LO stemming from the transmitter. For the highest pump power we measured 3.7 dB and 3.9 dB, respectively. The difference can be explained by phase noise which we determined by the fit to be 56 mrad for the real LO and 29 mrad for the regular one. The total transmission efficiency of the setup, estimated from the fit in Fig.~\ref{fig:results global}a is approximately $64\,\%$ which limited the measured squeezing to about 3.7\,dB.

We then added a 1, 5, 10 and 40\,km fiber between the squeezed light source and the receiver and measured squeezing with both the real LO from the receiver laser and the regular LO from the transmitter laser. The power of the pump field of the squeezed light source was set to 2.6\,mW. The results can be seen in Fig.~\ref{fig:results global}b. We note that for fair comparison we did not transmit the regular LO from the transmitter laser through the same fiber but instead used another fiber of constant length. This however let to the accumulation of phase drifts caused by the long fibers the squeezed light travelled through. While the coherence length of the lasers by far exceeds 40\,km, the phase fluctuations in the fiber made it necessary to use a fast EOM for phase locking. Assuming $0.18$\,dB/km attenuation coefficient of the optical fibers, as per manufacturer's specifications, using (\ref{eq:quadrature PSD}) we obtained the fitting curve in Fig.~\ref{fig:results global}b. While the squeezed quadrature variance at low and high distances is in good agreement with the theoretical model, there is a mismatch of $0.7$ dB (corresponding to 15\% deviation) at 5 km and 10 km. This is most likely explained by a drift in the temperature of our squeezer crystal, observed at times in the laboratory, due to an unstable temperature control that drives the pump field slightly off the phase-matching condition necessary to generate single-mode squeezing. The result would be a lower parametric gain of the squeezer at 5 km and 10 km compared to other distances. Besides this experimental imperfection, Fig.~\ref{fig:results global}b still shows the expected behavior of squeezing over distance, in relation to the fiber loss specifications.

The theoretical model of Fig.~\ref{fig:results global}b also takes into account the standard deviation of phase noise, estimated from a linear fit on the phase noise measurements of Fig.~\ref{fig:results global}c. The latter reports the estimated standard deviation of the pilot mode's phase noise under phase locking, at the given distances. The estimation procedure is described in detail in the supplementary material. The linear fit provided by this measurement and fed into the theoretical model of Fig.~\ref{fig:results global}b, resulted in a phase noise coefficient of 1.7 mrad/km for the real LO and 2.8 mrad/km for the regular LO. The increase in phase noise with transmission distance can be explained by two effects: Firstly, the signal-to-noise ratio of the pilot beam drops with transmission distance, thereby reducing the quality of the error signals at the input of the PI-controllers. Secondly, the phase fluctuations become stronger while the light travels through the fiber. Since the loop gain stays constant, the deviations from the lock point increase. Notably the phase noise at 40 km is higher for the regular LO than for the real LO. 

This effect can also be seen in Fig.~\ref{fig:results global}d where we investigated the squeezed light transmitted through 40\,km of fiber in more detail. Here we varied again the pump power of the squeezed light source and measured the noise variances of the squeezed and anti-squeezed quadratures using both LOs. The overall optical transmission was determined from the fit of the theoretical model of Eq.~\ref{eq:quadrature PSD} to $13\%$ for the real LO and $11\,\%$ for the regular one. The deviation may come from a change of fiber in-coupling or polarization drifts. For the phase noise standard deviation we obtained 147\,mrad and 115\,mrad for the real and regular LO, respectively. At 40 km distance the phase drifts seem to be better compensated by the phase lock of the real LO than the regular one. 
\begin{figure}
	\centering
	\begin{tabular}{c}
		\captionsetup{width=180pt}
		\subfloat[]{\includegraphics[scale=0.3]{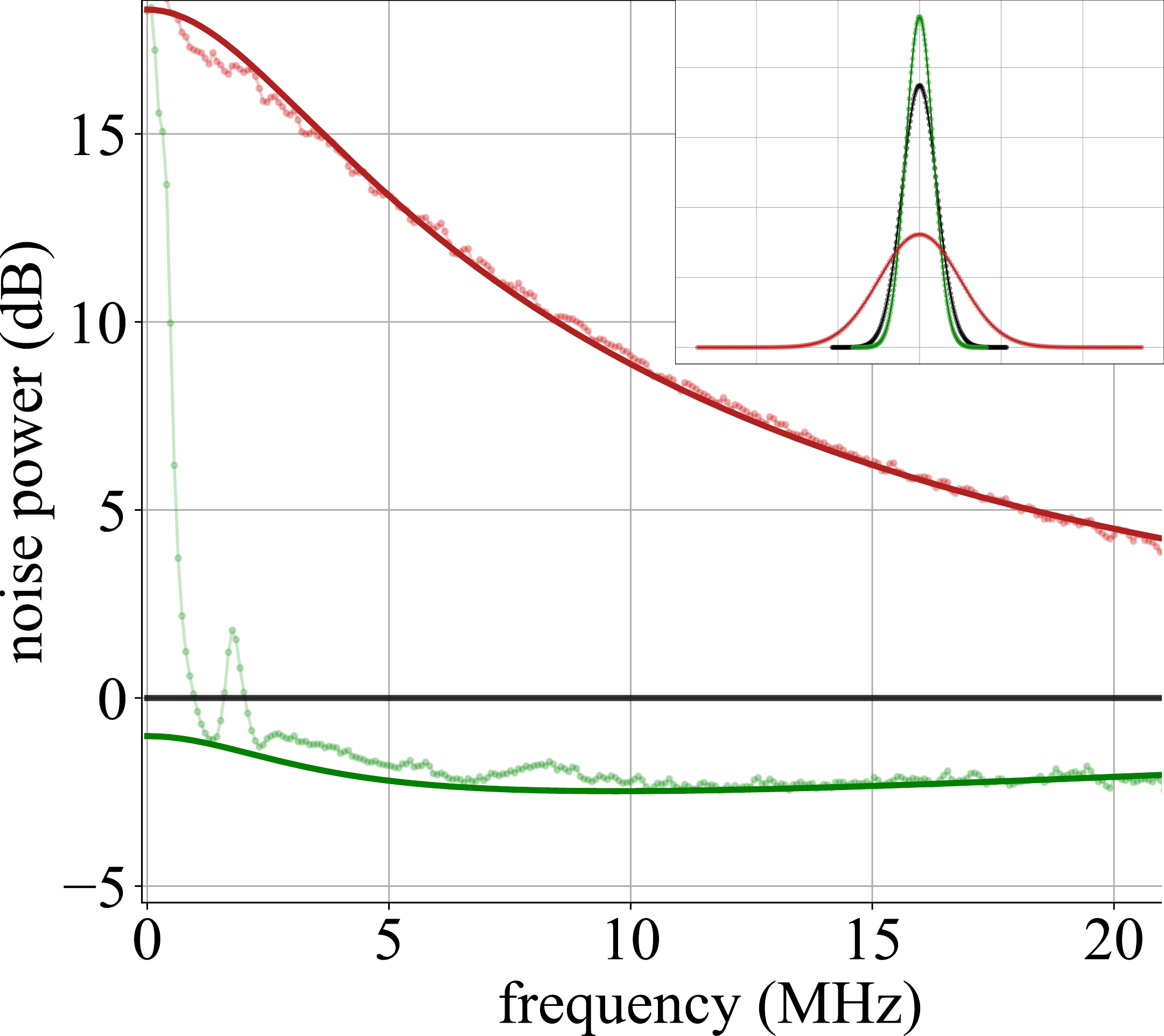}}    \\
		\begin{tabular}{cc}		         \subfloat[]{\includegraphics[scale=0.3]{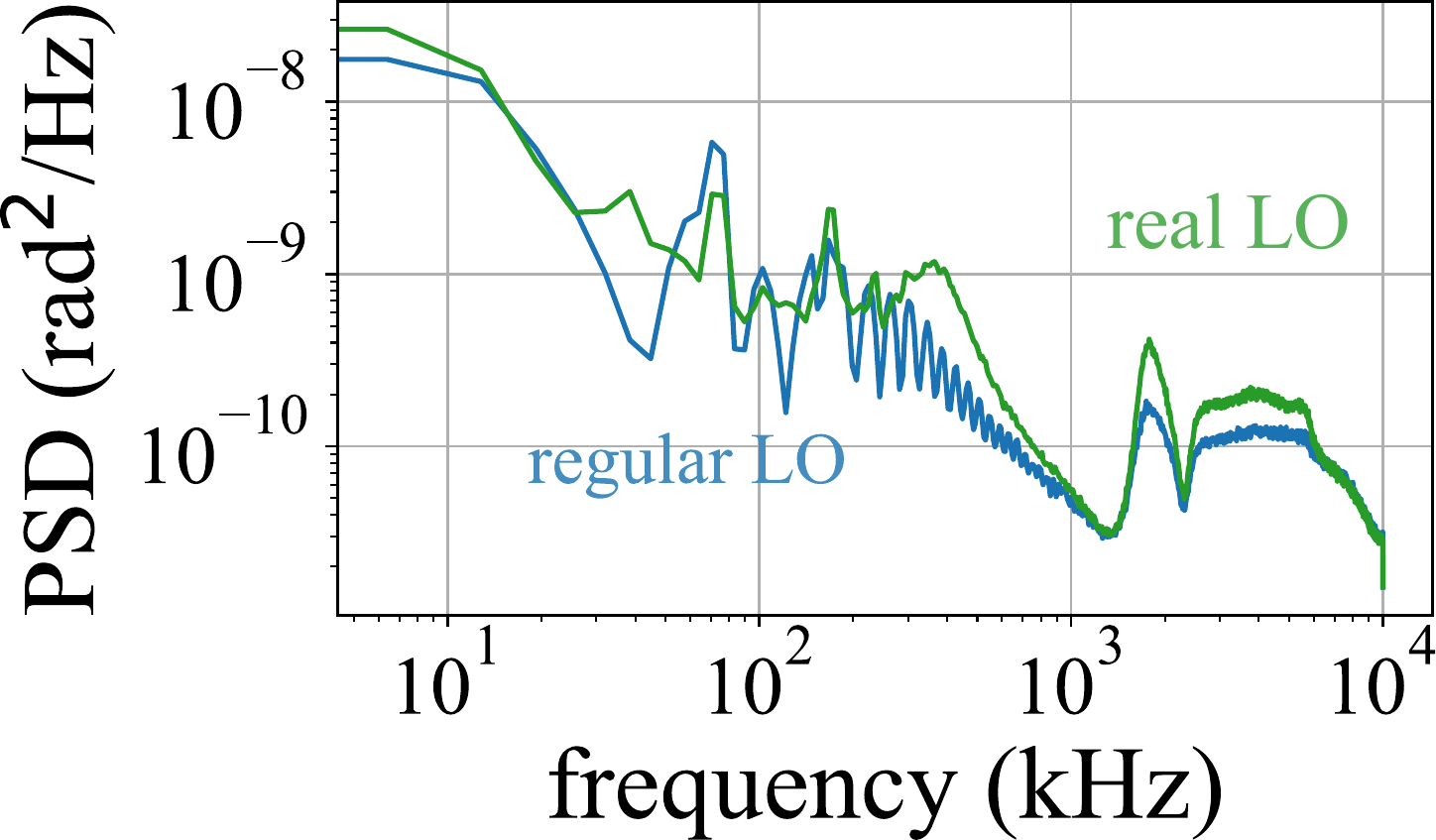}}  & \captionsetup{width=120pt}\subfloat[]{\includegraphics[scale=0.3]{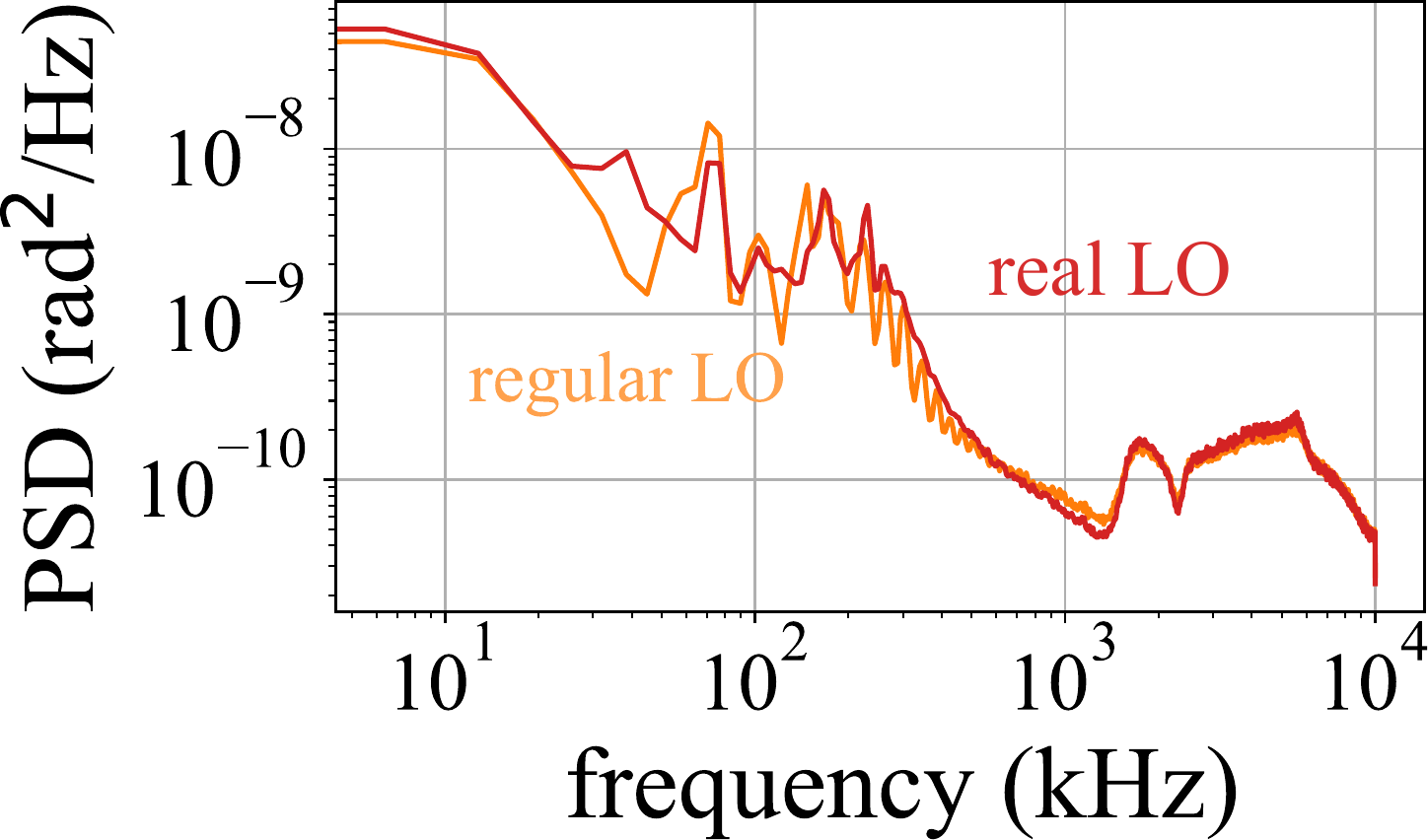}}\\
			\subfloat[]{\includegraphics[scale=0.3]{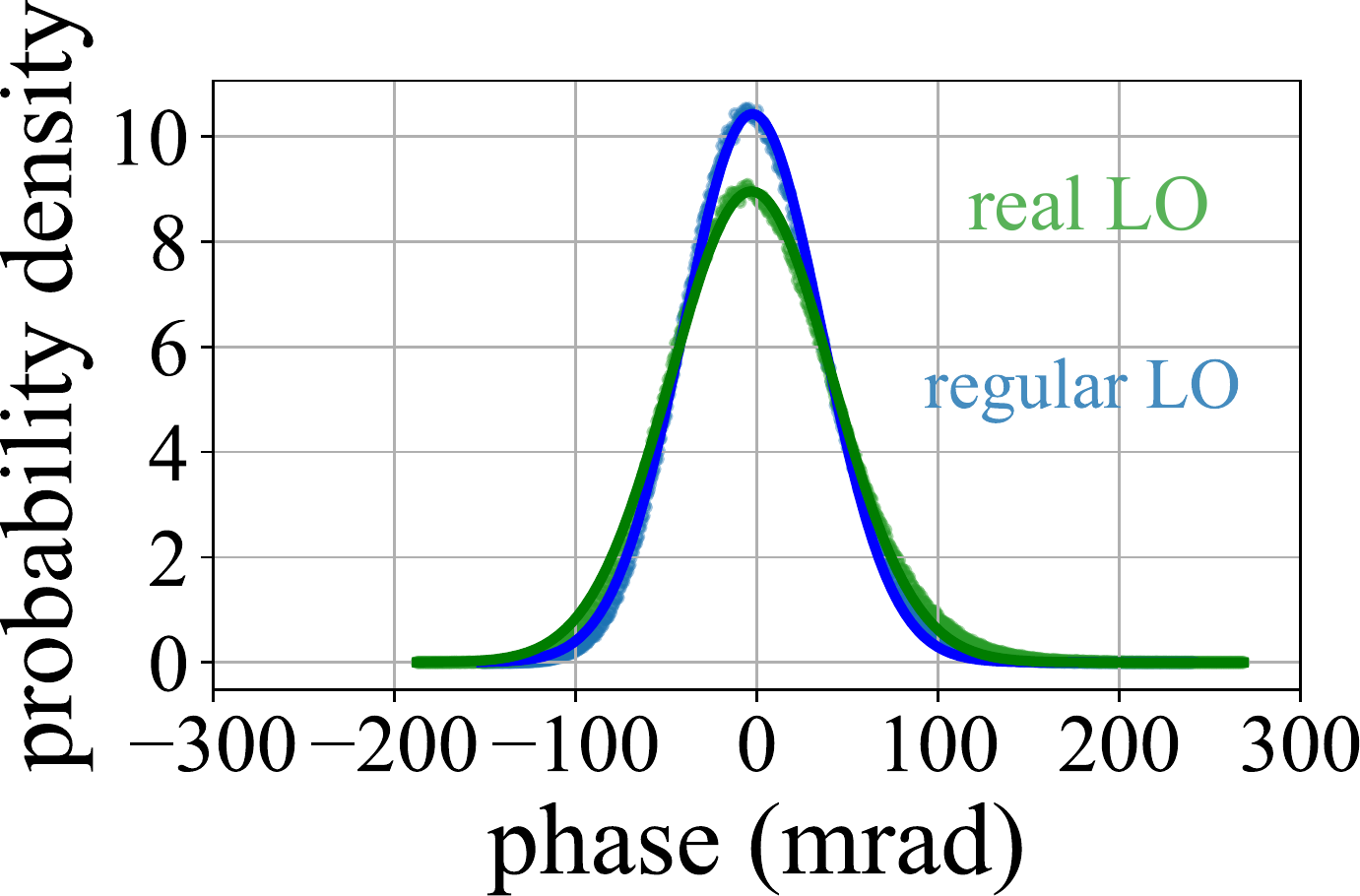}} & \subfloat[]{\includegraphics[scale=0.3]{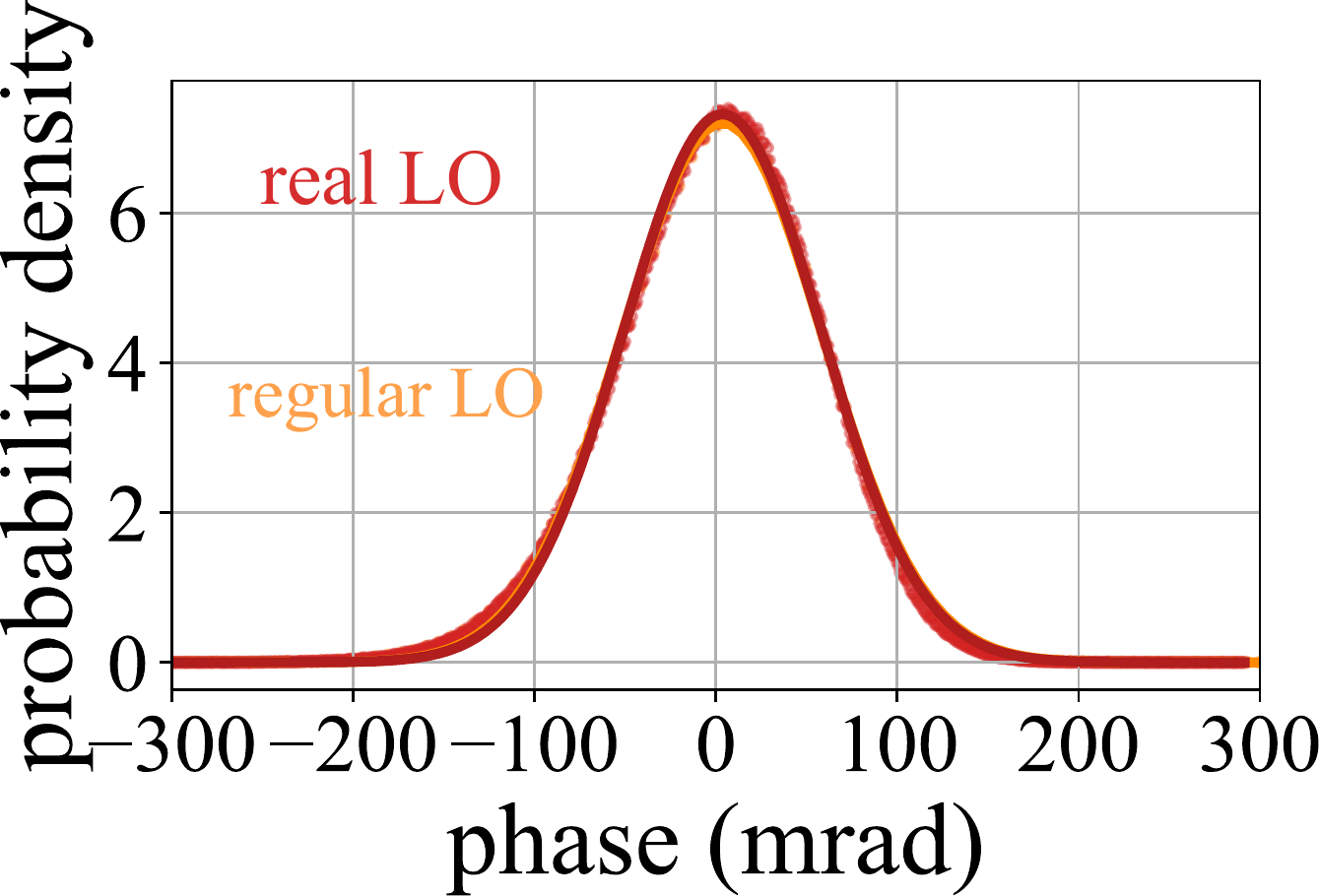}}
		\end{tabular}
	\end{tabular}
	\captionsetup{width=\textwidth}\caption{\small{(a) presents the quadrature measurements performed with the real local oscillator, for 5 km fiber length and 2.6 mW pump field optical power. The largest plot represents the statistical power of the squeezed (below zero) and antisqueezed (above zero) quadratures of the signal, relative to vacuum quadrature noise. As an insert, we include the calculated histogram-based quadrature probability density functions. Figures (b) and (c) represent the power spectral densities (PSD) of the estimated phase noise in the same measurements. Figures (d) and (e) show the histogram-based probability density functions of the phase noise together with Gaussian fits (solid lines). real local oscillator: squeezed quadrature in green, antisqueezed quadrature in red. Regular local oscillator: squeezed quadrature in blue, antisqueezed quadrature in orange. Black color represents the vacuum quadrature.}}
	\label{fig:results detail}
\end{figure}

At a distance of $5$\,km  we investigated the quadrature noise spectrum and phase noise in more detail. Figure~\ref{fig:results detail}a shows the quadrature noise power spectrum of the squeezed light in the range 0 to 20 MHz measured with the real local oscillator. The noise power has been normalized to the noise power of the vacuum state. The pump power was $2.6$\,mW. The fit of the theoretical model from Eq.~\ref{eq:quadrature PSD} reveals an optical transmission of $50.0\,\%$, corresponding to about 1.3 dB fiber attenuation, and a phase noise standard deviation of 65\,mrad. The insert shows histograms of the noise for the squeezed state in the 0 to 20 MHz frequency range in the squeezed and anti-squeezed quadrature and the vacuum state as comparison.

By acquiring the 40 MHz pilot signal with a data acquisition system digitizing the coherent receiver output we extracted the residual phase noise power spectral density. The phase estimation procedure, which is explained in detail in the supplementary material, has been performed for the regular and the phase-locked real local oscillator. Figure~\ref{fig:results detail}b shows the phase noise spectrum for the squeezed quadrature while Fig.~\ref{fig:results detail}c shows it for the antisqueezed quadrature and the corresponding phase histograms are shown in Fig.~\ref{fig:results detail}d and e, respectively. When phase locked, the signal-to-noise ratio of the pilot is different in the two cases due to the antisqueezing which lowers it by around 5 dB, however, the performance of the phase lock seems to be similar.At low frequencies the phase noise of the real LO is slightly larger than the phase noise of the regular LO. The phase histograms are taken over the entire frequency range. Here the peak around 80 kHz in the spectra for the regular LO seems to compensate the better phase noise behaviour at low frequencies.

\section{Discussion}
\label{Discussion}
The developed system allowed homodyne detection of up to 3.7 dB squeezed states of light over optical fiber spools of up to 40 km length. The control system exhibited a minimum phase noise standard deviation of $(43.78 \pm 0.07)$ mrad, corresponding to $(2.508 \pm 0.004)^\circ$, under phase control. In particular, it enables the measurement of squeezed states and other quantum states of light spatially separated from the source without the distribution of high-power local oscillators through the fiber network (thereby avoiding detrimental effects such as non-linear Brillouin scattering in optical fibers \cite{yaman_guided_2021}).
The maximum measured amount of squeezing was not primarily limited by phase noise, but rather by the optical transmission efficiency of the squeezed mode through the experimental setup, which can be improved by accurate single-mode fiber coupling and by optimizing the detection efficiency in future implementations. The phase noise performance of the phase control system could be improved by reducing the electronic amplification noise at the output of the homodyne detector and at the input of the control actuators, and by performing the downmixing and low-pass filtering that yield the signals $i(t)$ and $q(t)$ inside the FPGA rather than at its input, thereby avoiding the use of potentially noisy analog mixers and low-pass filters. Furthermore, the three employed PI control systems should be carefully designed, to ensure that phase noise is effectively compensated for over its full Fourier spectrum support, and to avoid detrimental interference between the different control systems in action. 
\section{Conclusions}
\label{Conclusions}
This work has demonstrated the possibility of transmission and homodyne detection of generalized quadratures of squeezed states of light at 1550 nm wavelength over at least 40 km single-mode optical fiber (compatible with large metropolitan networks), using a phase-locked real local oscillator. Our system makes use of standard telecom components, such as piezoelectric fiber stretchers and electro-optic modulators, and widely used general-purpose digital signal processing devices, such as FPGAs. It is a general homodyne detection technique that allows for the characterization of arbitrary states over long distances in a network, as needed for a variety of quantum information tasks. For quantum key distribution, using a true local oscillator disables eavesdropping attacks that target the transmitted local oscillator \cite{huang_quantum_2013}, and using a real-time phase control avoids the need for accurate and computationally heavy phase estimation techniques, often performed in post-processing within current proof-of-concept CVQKD experiments \cite{zhao_phase_2019}. The system becomes particularly interesting if applied to the quadrature detection of non-Gaussian and entangled states of light in distributed  photonic quantum computing based on measurement-induced quantum computing modules \cite{larsen_deterministic_2021}, in distributed quantum sensing systems \cite{guo_distributed_2020, zhang_distributed_2021} and in quantum communication networks \cite{jouguet_experimental_2013, yin_entanglement-based_2020}. The quadrature detection of non-classical light with real local oscillators therefore constitutes an important technique in future quantum photonics networks for quantum computing, quantum sensing and quantum communication.
\section{Acknowledgements}
\label{Acknowledgements}
The authors acknowledge support from European Union’s Horizon 2020 research and innovation program CiViQ under grant agreement no.\ 820466, from the Independent Research Fund Denmark (Sapere Aude Starting Grant, grant agreement no.\ 0171-00055B), from the QuantERA grant ShoQC through the Innovation Fund Denmark (case no. 9085-00011B), and from the Danish National Research Foundation, Center for Macroscopic Quantum States (bigQ, DNRF142).
\section{Data availability} Data underlying the results presented in this paper are not publicly available at this time but may be obtained from the authors upon reasonable request.	
\section{Disclosures}
The authors declare no conflicts of interest.
\bibliographystyle{OSA}
\bibliography{references}
\end{document}